\documentclass[aip,rsi,reprint,graphicx]{revtex4-1}
\usepackage[pdftex]{graphicx}
\usepackage{graphicx}
\usepackage{color}
\draft 
\preprint{AIP/123-QED}
\usepackage{amssymb}

\begin{document}
\title{Onset of lasing in small devices: the identification of the first threshold}

\author{T. Wang}
 \email{wangtao@hdu.edu.cn} 
 \affiliation{Universit\'e C\^ote d'Azur, Institut de Physique de Nice (INPHYNI), CNRS UMR 7010, Nice, France}
 \affiliation{School of Electronics and Information, Hangzhou Dianzi University, Hangzhou, China}

\author{G.P. Puccioni}
 \affiliation{Istituto dei Sistemi Complessi, CNR, Via Madonna del Piano 10, I-50019 Sesto Fiorentino, Italy}
 
\author{G.L. Lippi}
 \affiliation{Universit\'e C\^ote d'Azur, Institut de Physique de Nice (INPHYNI), CNRS UMR 7010, Nice, France}

\date{\today}

\begin{abstract}
We present a simple and flexible technique for identifying the onset of coherent emission in lasers, from the meso- to the nano-scale, which makes use of photon counting and a {\it small amplitude} modulation added to the pump. The optimal modulation frequency is obtained from the radiofrequency power spectrum of the unperturbed laser emission.  The identification of the lasing onset rests on the appearence of a resonance in the response of the zero-order autocorrelation function ($g^{(2)}(0)$) plotted as a function of the pump rate. The intrinsic simplicity of this technique and its use of photon counting make it an excellent tool for certifying the onset of laser emission in nanoscale sources.
\end{abstract}

\maketitle

\section{Introduction}
The question of threshold identification in small lasers has plagued the community since the inception of the first microlaser in the 1980s:  the Vertical Cavity Surface Emitting Laser (VCSEL)~\cite{Soda1979}.  A tremendous amount of work has been devoted both to the conceptual clarification of the threshold issue, which becomes elusive as the cavity size shrinks, as well as to its practical measurement~\cite{Ma2013, Ning2013}.  Indeed, as is well-known, the laser response becomes progressively smoother as the cavity shrinks and the characteristic jump which identifies the threshold in macroscopic lasers disappears to give rise to a gradually growing output power~\cite{Yokoyama1989,Bjork1991}.  

Current common wisdom loosely divides the laser response into three regimes (cf. for instance~\cite{Pan2016}):  a first one dominated by spontaneous emission (lowest pump values), a second one where amplified spontaneous emission (ASE) controls the laser output (intermediate values, more or less corresponding to the region with the largest slope in the output) and the truly lasing regime (the ``upper branch" of the emission curve).  This classification is qualitative and, although conceptually valid, is too vague for any practical use especially because devoid of quantitative criteria for the regime demarcation and for coherence definition.  One interesting and ingenious step forward has been recently accomplished in the identification of the pump value for which the laser field acquires coherence~\cite{Pan2016}.  This technique applies, however, only to nanolasers which are optically pumped by laser pulses with duration several nanoseconds -- a somewhat unusual pump scheme which in the long run will not be suitable for most practical applications~\cite{Service2009}. Since, in addition, the aim is to move towards continuously pumped devices -- as proven~\cite{Lu2012,Gu2014,Lu2014,Hayenga2016,Zhu2016,Molina2016}, and reviewed~\cite{Ding2012,Gwo2016} -- this scheme~\cite{Pan2016} does not offer a general-purpose solution for threshold identification.  The strong push for a correct identification of threshold in very small devices, in support of the claim of laser emission, justifies the recent directed efforts~\cite{nature2017}.

The very low photon flux emitted by a nanolaser imposes~\cite{Pan2016} the use of the only technique which is sufficiently sensitive for detection:  photon counting.  Thus, successful threshold identification must rely on photon counting and simple but reliable manipulations of the information that it provides.  In the following, we present a method to identify the onset of coherent emission in small-scale lasers based on the zero-delay second-order autocorrelation of the field intensity:  $g^{(2)}(0)$.  We classify the laser size on the basis of the expected fluctuations at threshold~\cite{Rice1994} and identify nanolasers in the range $10^{-2} \lessapprox \beta \lessapprox 1$ and mesoscale lasers $10^{-4} \lessapprox \beta \lessapprox 10^{-2}$, where $\beta$ represents the fraction of spontaneous emission coupled into the lasing mode. We have previously shown that mesoscale lasers present features which are comparable, even though less extreme, to those of nanolasers~\cite{Wang2015,Wang2016b,Puccioni2015} and that therefore measurements performed in mesoscale devices can be transferred to the nanoscale.  

First, we will present experimental proof for our method based on measurements performed on a mesoscale laser, then we shall give theoretical support for transferring this technque to the nanoscale. The central point of the procedure rests on the application of a small-amplitude modulation to the biased laser and in the interpretation of the {\it resonance} which appears in the shape of $g^{(2)}(0)$ as a function of pump current.  The smallness of the modulation amplitude is at the same time a strict requirement, as we will later see, but also a very fortunate technical coincidence, since weak radiofrequency modulation signals are easier to handle.

\section{Experimental Setup}
The experimental setup is schematically shown in Fig.~\ref{setup} (details~\cite{Wang2015} in the Supplementary Information). The experiment is performed on an electrically-pumped, small-diameter ($d \approx 6 \mu m$) VCSEL operating at $\lambda = 980 nm$ (Thorlabs, model VCSEL-980), capable of emitting a maximum output power $P_{max} \approx 1.85 mW$  for a pump current $i = 10 mA$ and mounted on a TEC module (Thorlabs TCLDM9).  The laser is supplied by a stabilized, high-resolution ($1 \mu A$, accuracy $\pm 20 \mu A$) current source (Thorlabs LDC200VCSEL) and is kept at constant temperature ($T = (25.0 \pm 0.1)^{\circ}C)$) thanks to home-built apparatus.  The pump modulation is added through the modulation input of the TEC module and is obtained from a sinusoidal function generator (E4421B, Hewlett).  As shown in Fig.~\ref{setup}(a), the laser radiation is coupled into a multimode fiber after passing through a Faraday isolator (QIOPTIQ 8450-103-600-4-FI-980-SC) to be detected by a fast, amplified photodetector (Thorlabs PDA8GS) with $9.5 GHz$ analog bandwidth.  The data is acquired through a LeCroy Wave Master 8600A oscilloscope with $6 GHz$ analog bandwidth (sampling time $\Delta t = 0.1 ns$) and up to $5 \times 10^6$ samples in each trace.  At variance with the scheme commonly used for nanolasers (quantum coincidence measurements), we reconstruct the zero-order autocorrelation from the time trace of the detected signal.  This does not in any way impact on the generality of the technique, since it is based on the knowledge of $g^{(2)}(0)$, independently on the measurement details.

\begin{figure}
\centering
\includegraphics[width=1\linewidth,clip=true]{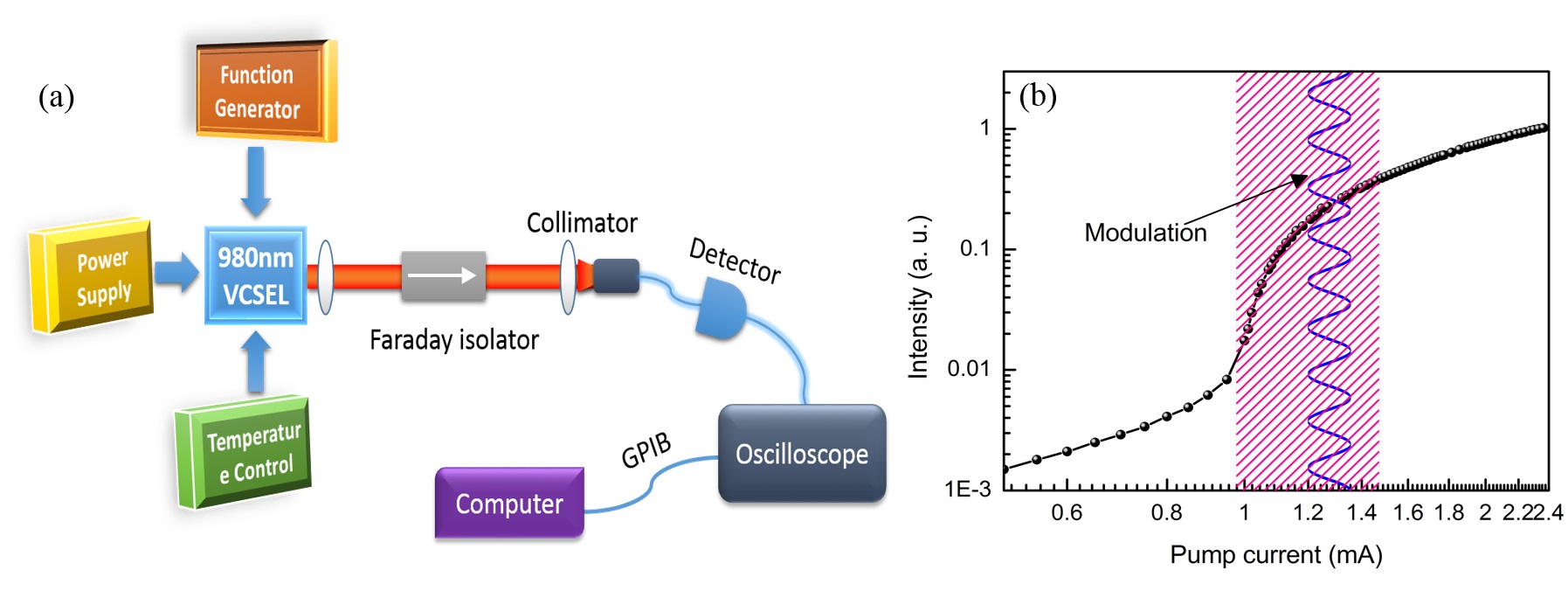}
\caption{(a)Experimental setup;(b)Input-output laser response with superposed investigated region (shaded) and sinusoidal modulation.}
\label{setup}
\end{figure}

\section{Results and Discussion}
The input-output lasing characteristic function curve in log-log scale is shown in Fig. ~\ref{setup}(b), which displays an ``S-slope'' transition between spontaneous and stimulated emission typical of microscale lasers. The ``kink'' point of the slope curve can be easily recognized at 0.95 mA, which traditionally corresponds to the so-called ``threshold'' point. The estimated spontaneous emission coupling factor for this laser ~\cite{Wang2015} is $\beta \approx 10^{-4}$.  The shaded area ($0.95 mA \le i \le 1.4 mA$) highlights the main region of interest for this work.  

\begin{figure}
\centering
\includegraphics[width=0.65\linewidth,clip=true]{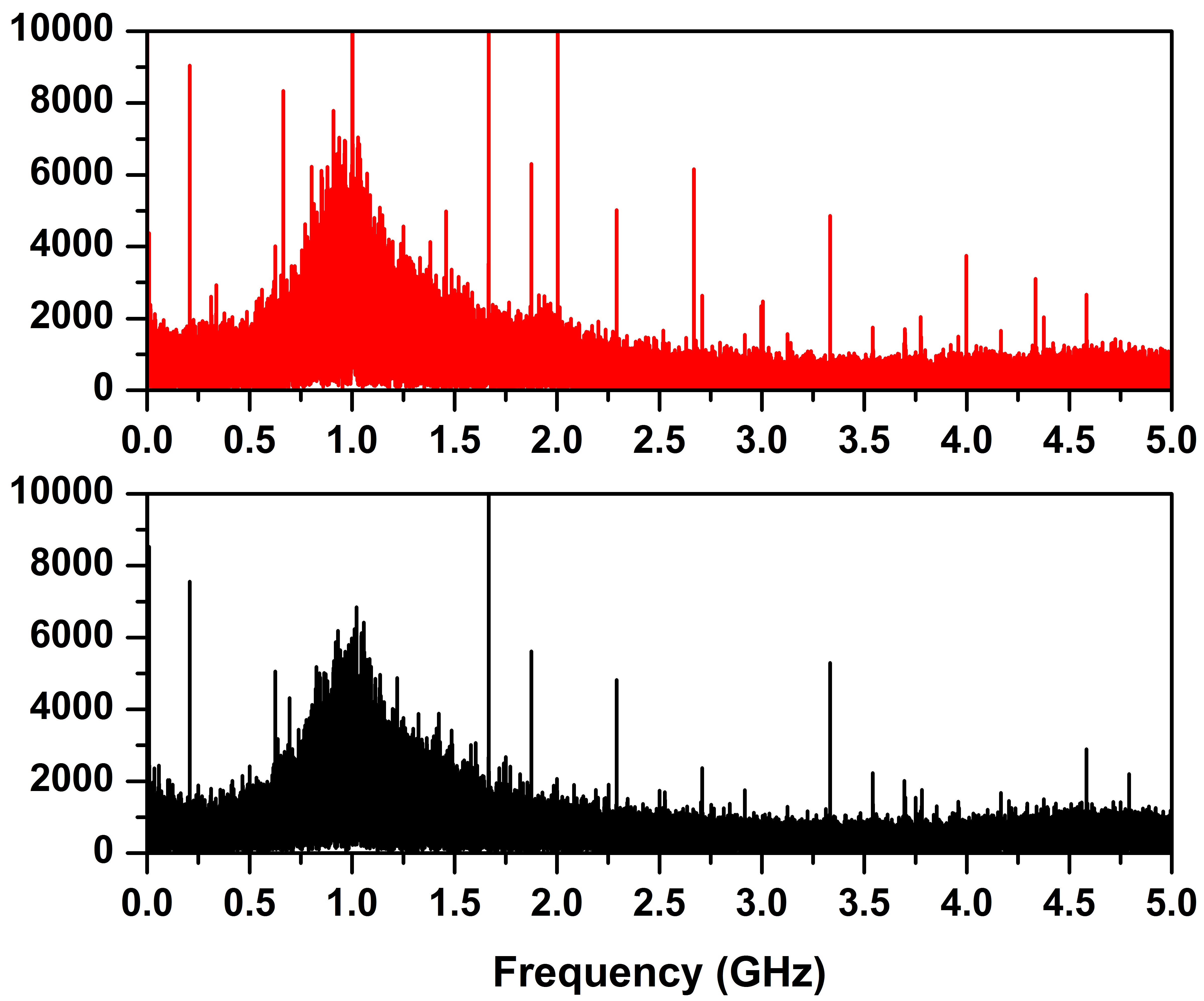}
\caption{Laser rf spectra at bias $i = 1.06 mA$ without (bottom) and with a small-amplitude modulation (top)}
\label{mAspectrum}
\end{figure}

Fig.\ref{mAspectrum} shows the laser's radiofrequency (rf) spectra, obtained by Fourier-transforming the temporal data traces, with (red, top panel) and without (black, bottom panel) small amplitude modulation. The modulation frequency is $1 GHz$ (sharp red line), located on top of the broad spontaneous laser resonance.  Harmonics of the modulation are visible at $2$, $3$, and $4 GHz$ while around $2 GHz$ an additional broader feature is seen to grow out of the laser spectrum (red).  The additional peaks present in both panels correspond to background noise.

\begin{figure}
\centering
\includegraphics[width=0.65\linewidth,clip=true]{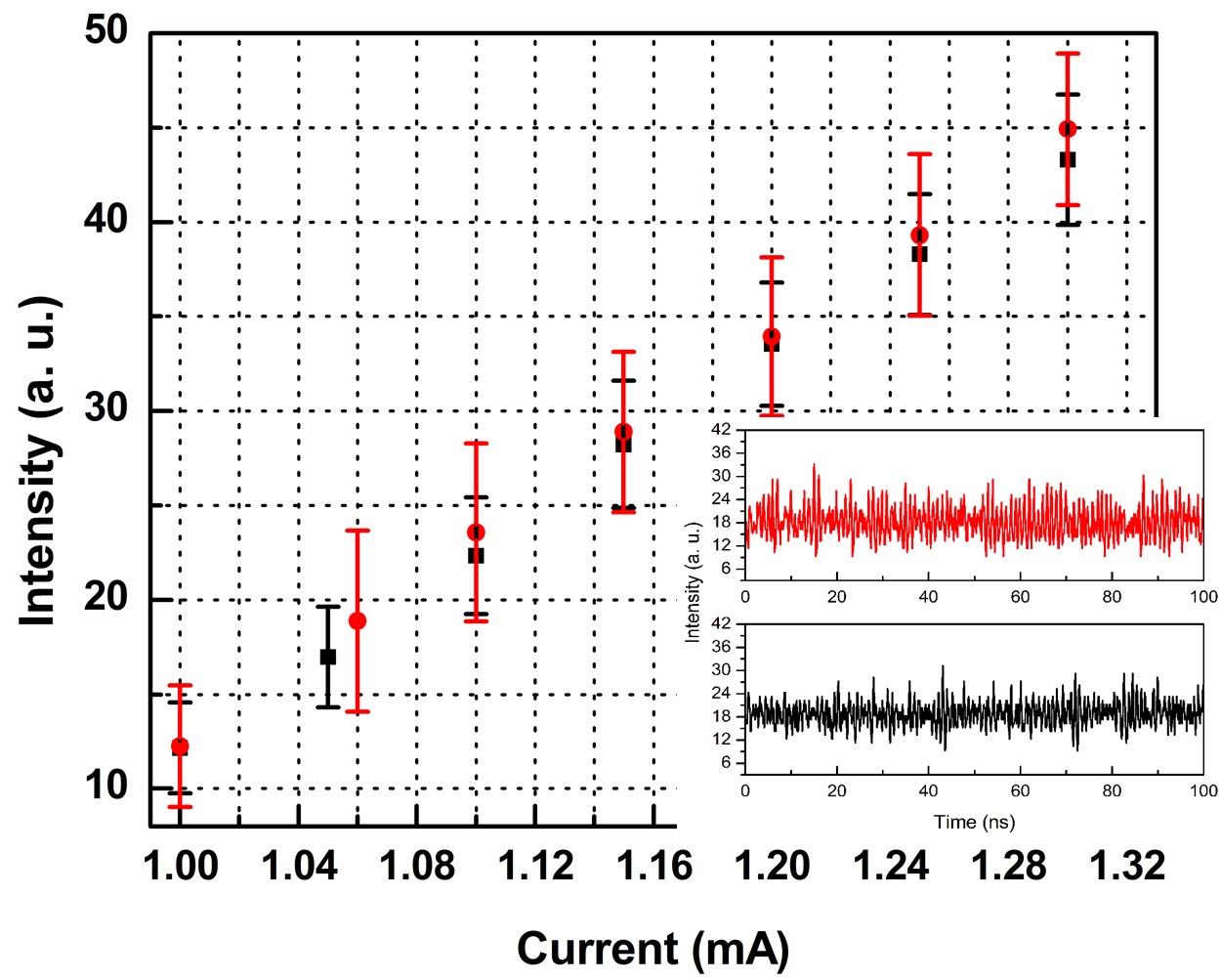}
\caption{Average laser intensity (symbols) and standard deviation (error bars) as a function of bias pumping current in the absence (squares, black) and in the presence (dots, red) of a small-amplitude modulation. The inset shows two typical temporal traces of the intensity at bias current i = 1.06mA in the presence (top) and in the absence (bottom) of the external modulation.
}
\label{SSM2}
\end{figure}

Fig.\ref{SSM2} shows the average intensity values (dots), with standard deviation (error bars), in the absence (squares, black online) and in the presence (circles, red online) of the small amplitude modulation for different pump values in the threshold region, plotted in linear scale (the apparent nearly linear relationship follows from the small interval of current values):  the average power remains (nearly) unchanged, while its standard deviation is amplified by the presence of the modulation (red bars).  The amplification is maximal at $i = 1.06 mA$ (compare red and black error bars), for which a typical temporal trace is shown in the inset of Fig.~\ref{SSM2}:  the oscillation is amplified by the small-amplitude modulation (red, top curve). Notice that the difference between the amplitude fluctuations reduces as the bias current is increased (upwards of $i = 1.2 mA$), while in this range the average output value is clearly raised by the presence of the modulation.

\begin{figure}
\centering
\includegraphics[width=0.65\linewidth,clip=true]{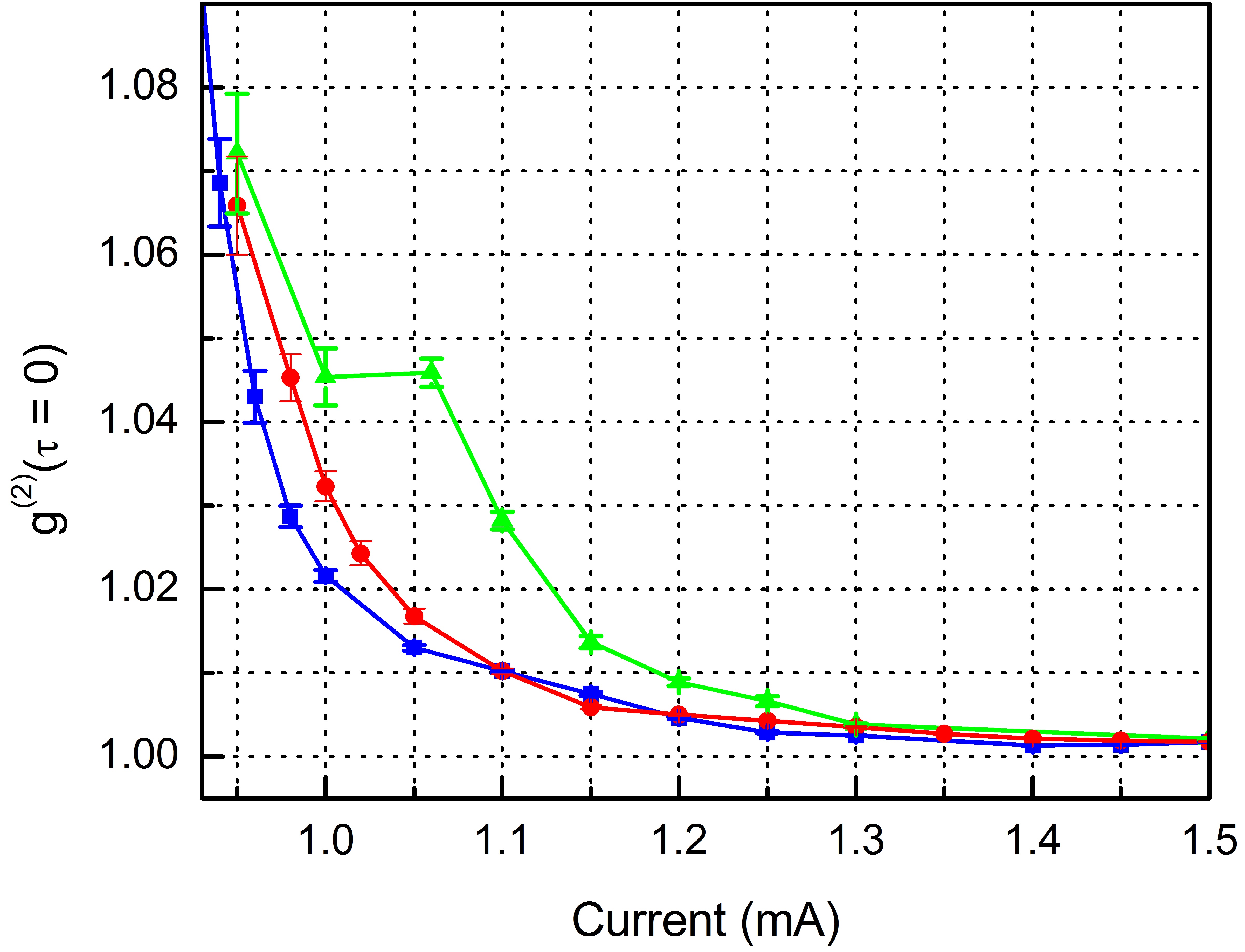}
\caption{Second-order autocorrelation functions for the laser under free running (dots, red online), small signal modulation (triangles, green online -- modulation amplitude 5\%) and large amplitude modulation (squares, blue online -- modulation amplitude 30\%).}
\label{Mcorre}
\end{figure}

In order to obtain a quantitative identification of the modifications induced by the modulation, we compute, from the digitized intensity values, the zero-delay second-order autocorrelation function $g^{(2)}(\tau = 0)$ both with and without modulation,
\begin{equation}
g^{(2)}(\tau = 0) \equiv \frac{\langle I^2 \rangle}{{\langle I \rangle}^2} \, ,
\end{equation}
where $I$ represents the measured intensity, $\langle \cdot \rangle$ the average operation.  $g^{(2)}(0) = 1$ denotes fully developed field coherence (Poisson limit).  The dependence of $g^{(2)}(0)$ on the bias current for a fixed, small modulation amplitude, is plotted in Fig.\ref{Mcorre} using ten temporal intensity traces to compute average and standard deviation.

In the absence of modulation (dots, red online, Fig.~\ref{Mcorre}) $g^{(2)}(\tau = 0)$ displays the same features already observed in lasers of the same kind~\cite{Wang2015,arXiv}, characterized by a rapid decay of the autocorrelation until $i \approx 1.05 mA$, followed by a slower decay with eventual convergence to the Poisson limit.  The first decay has been identified as a regime of independent pulses~\cite{Wang2015}, followed by irregular oscillations which gradually diminish until the true lasing regime is attained.

In the presence of a small-amplitude modulation (triangles, green online, Fig.~\ref{Mcorre}) a very obvious plateau appears in the shape of $g^{(2)}(0)$ in the interval $1.0 mA \lessapprox i \lessapprox 1.06 mA$, followed by a renewal of the decay towards the autocorrelation values obtained in the absence of modulation (circles, red online).  The presence of the plateau signals the ceasing of the coherence growth in the corresponding bias current interval. Here it is important to understand the origin of such drastic change in behaviour, since this is a strong indicator of a change in the laser dynamics.

Meso- and nanolasers are all semiconductor-based devices and as such are characterized by a carrier dynamics slower than the photon dynamics (Class B lasers~\cite{Tredicce1985}).  This implies that the best-known properties of photon statistics~\cite{Loudon1982,Mandel1995,Arecchi1971} do not apply~\cite{arXiv} and that the evolution of coherence follows a more complex path.  In particular, coherent emission starts with the coupled dynamics of carriers and photons in a {\it noisy} way due to the interplay of the time constants~\cite{Narducci1988,Lugiato2015} for carriers and photons, traditionally known in semiconductor lasers under the name Relaxation Oscillations~\cite{Coldren2012}.  The particularly noisy nature of the oscillation is due to the discreteness of the problem:  since the photon and carrier numbers are rather small in the threshold region in lasers with ``large" $\beta$-values~\cite{Rice1994}, the discreteness reflects on the dynamics of the system by rendering it much noisier ~\cite{Lebreton2013,Puccioni2015,Wang2016b,Wang2016} than in macroscopic lasers ($\beta < 10^{-5}$). It is important to realize that the appearence of oscillations, even though irregular, signals the first onset of coherence, since such oscillations can only occur in the presence of a coupled dynamics between {\bf \em carriers} and the {\bf \em coherent component} of the electromagnetic field.  Thus, we can consider the corresponding pump value as the point at which coherent emission starts, even though in a {\it noisy} kind of fashion.  The dynamics which precedes the coherent oscillation instead is due to the amplification of spontaneous emission in coherent but independent bursts~\cite{Wang2017}, which do not carry phase information (thus coherence) from one pulse to the next:  this is the regime commonly called {\it Amplified Spontaneous Emission} which, contrary to common belief, does not extend over the whole nonlinear response region.  More details on these points will be given in a forthcoming publication.

The blue squares in Fig.\ref{Mcorre} show the result of applying a ``large" modulation amplitude ($\approx 30\%$ of the pump value) to the laser in place of a small one:  The resonance in $g^{(2)}(0)$ has entirely disappeared!  This is easily understood since the characteristic frequency of the Relaxation Oscillations, though noisy and therefore broadened, depends on the bias value.  If the modulation amplitude is large, the effective bias changes considerably from one pump value to the next, thereby continually shifting the resonance frequency and destroying the laser's response to modulation. 

\begin{figure}
  \includegraphics[width=0.65\linewidth,clip=true]{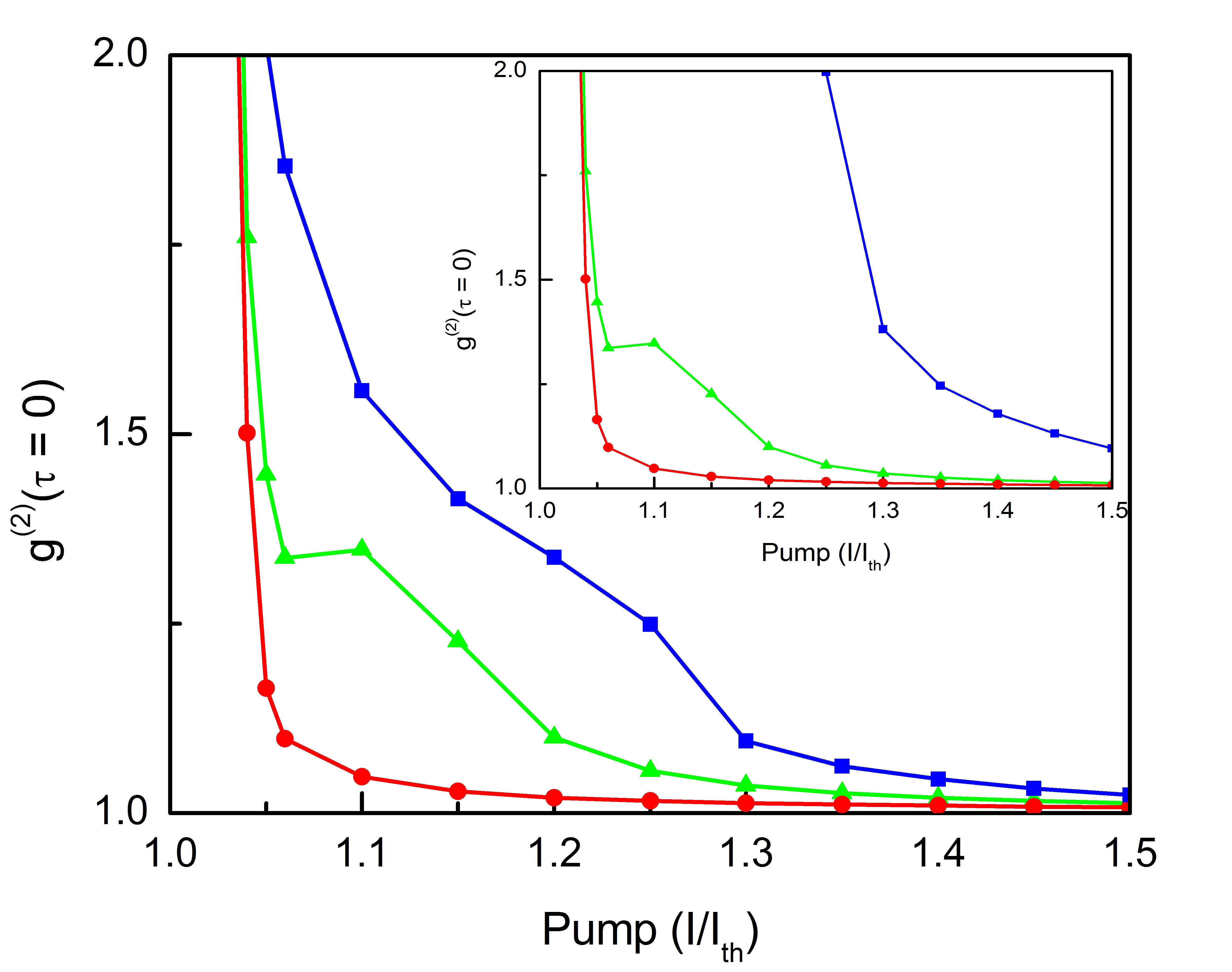}
  \caption{Numerical simulation results, obtained from the Stochastic Simulator~\cite{Puccioni2015}, of second-order autocorrelation functions for the laser under free running (dots, red online), small signal (triangles, green online, $\pm 2.5 \%$ modulation amplitude) and large signal modulation (squares, blue online, $\pm 30 \%$ modulation amplitude).  The curves are rescaled to have the same peak in the autocorrelation at $\frac{I}{I_{th}} = 1$, inset.  In the main panel the large modulation curve is rescaled to show that no resonance occurs around $\frac{I}{I_{th}} \approx 1.1$.}
  \label{AMF}
\end{figure}

Theoretical support to these findings, and proof for extending them to lasers smaller than the one used in the experiment, is obtained through the use of a Stochastic Simulator~\cite{Puccioni2015}, which allows for the investigation of the fully stochastic dynamics in the threshold region for lasers of any size.  Fig.\ref{AMF} shows the predicted value of the autocorrelation for $\beta = 10^{-4}$, as in the experiment (cf. figure caption for details).  The shape of $g^{(2)}(0)$ is shown in the absence of modulation (dots, red online), for a small amplitude modulation (triangles, green online), and for large-amplitude modulation (squares, blue online).  Notice that the values of $g^{(2)}(0)$ are much larger than in the experiment since in the latter there is a strong filtering action due to the bandpass of the detection system~\cite{Wang2015}.  In addition, the numerical system is much more sensitive to the external modulation with large amplitude and the corresponding curve (squares, blue online) converges towards the Poisson limit at larger pump values than in the experiment.  The important point to be retained is that a {\it resonance} between the modulation and the autocorrelation does not appear at large modulation -- a point easily understandable, as already mentioned, since the exploration of a large interval of pump values (through large-amplitude modulation) washes out any possible match between external signal and internal relaxation oscillation frequency, which possesses a clear dependence on the pump current~\cite{Coldren2012}.

Concentrating on the small-signal modulation (dots, red online) the match between predictions and observations is excellent:  the resonance appears very clearly for a pump value approximately 10\% above the nominal threshold (marked as $I_{th}$ in the figure label) which corresponds to the 12\% value from the experiment (taking $i = 0.95 mA$ as the conventional, nominal threshold).   In addition, this occurs only when the modulation amplitude is small, as observed in the experiment.

The choice of modulation frequency for the numerical simulations has followed the same criterion as for the experiment:  (a) the power spectrum of the free-running laser is computed and examined, and (b) the frequency corresponding to the (broad) peak of the spectrum is selected for the modulation.  Notice that, as in the experiment, the power spectrum develops a peak only when the bias current (pump) is  ``sufficiently large" (i.e., in the vicinity of what in the end turns out to be the estimated laser threshold).  Since the spectral peak displacement is slow, in pump, it is sufficient to examine a few spectra to obtain a satisfactory estimate of the modulation frequency.  

The simulations shown in Fig.~\ref{AMF} are run for a modulation frequency $f_{num} = 0.8 GHz$, close to the experimental one.  The discrepancy is well understood by the fact that the parameters we use in the simulation are indirect estimates (Supplementary Information~\cite{Wang2015}) and by the choice of model~\cite{Puccioni2015}.  The latter does not include the details of the physical description of lasing in semiconductor-based devices:  the qualitative, but close, similarity between the experimental results shows that the {\it resonance phenomenon} detected by the autocorrelation function (specifically $g^{(2)}(0)$) is general for all kinds of lasers (thus, could be applied for instance to solid-state microcavities) rather than being specifically related to semiconductor physics.

While we experimentaly prove the capabilities of this technique on a $\beta \approx 10^{-4}$ laser, confirming them with the numerical simulations, we can extend the prediction through the use of the same model~\cite{Puccioni2015}.  Running the same protocol (computation of $g^{(2)}(0)$ at a frequency identified by the first peak in the power spectrum) on lasers with growing $\beta$ values we find that the resonance persists until $\beta \approx 0.1$, i.e., well into the nanolaser scales, even though the {\it resonance step} in the shape of $g^{(2)}(0)$ becomes gradually less pronounced.  In a forthcoming paper we will give some explanations of the reasons for the reduced sensitivity to the laser to an external modulation.

\section{Conclusions}
In summary, we have shown an experimental technique for the identification of the onset of coherent emission in small-size lasers down to the nanoscale, based on the only measurement method (autocorrelation of counted photons) sufficiently sensitive to obtain quantitative information from such small sources.  This technique applies to any electrically or optically cw-pumped laser, as long as a small amplitude modulation can be added to the pump.  The modulation frequency is simply determined by taking rf spectra of the laser output in the absence of modulation and is therefore easily identified.  Furthermore, there is a good tolerance in the choice of modulation frequency since small lasers present rather large resonances in the threshold region.  The simplicity, flexibility and reliance of our method on the most sensitive optical detection technique -- directly applicable to any lasers through coincidence measurements -- promises to render it a most versatile tool for a clear identification of the onset of coherent emission in very small sources, possibly empowering the ``certification" of lasing action~\cite{nature2017}.  

\section{Acknowledgements}
We are grateful to A. Beveratos and I. Robert-Philip for discussions and to B. Garbin, F. Gustave and M. Marconi for experimental assistance.  Technical support from J.-C. Bery (mechanics) and from J.-C. Bernard and A. Dusaucy (electronics) is gratefully acknowledged.  T.W. acknowledges Ph.D. thesis funding from the Conseil R\'egional PACA and support from BBright.

\end{document}